\documentstyle[12pt,graphicx]{article}
\begin{document}
\baselineskip 0.6cm

\def\simgt{\mathrel{\lower2.5pt\vbox{\lineskip=0pt\baselineskip=0pt
           \hbox{$>$}\hbox{$\sim$}}}}
\def\simlt{\mathrel{\lower2.5pt\vbox{\lineskip=0pt\baselineskip=0pt
           \hbox{$<$}\hbox{$\sim$}}}}
\newcommand{\vev}[1]{ \langle {#1} \rangle }

\begin{titlepage}

\begin{flushright}
UCB-PTH-06/20 \\
LBNL-61989
\end{flushright}

\vskip 2.0cm

\begin{center}

{\Large \bf 
Predictive Supersymmetry from Criticality
}

\vskip 1.0cm

{\large
Yasunori Nomura and David Poland
}

\vskip 0.4cm

{\it Department of Physics, University of California,
           Berkeley, CA 94720} \\
{\it Theoretical Physics Group, Lawrence Berkeley National Laboratory,
           Berkeley, CA 94720} \\

\vskip 1.2cm

\abstract{Motivated by the absence of any direct signal of new physics 
so far, we present a simple supersymmetric model in which the up-type 
Higgs mass-squared parameter $m_{H_u}^2$ crosses zero at a scale 
close to the weak scale.  Such a theory may be motivated either by 
the conventional naturalness picture or by the landscape picture with 
certain assumptions on prior probability distributions of parameters. 
The model arises from a simple higher dimensional setup in which the 
gauge and Higgs fields propagate in the bulk while the matter fields 
are on a brane.  The soft supersymmetry breaking parameters receive 
contributions from both moduli and anomaly mediations, and their weak 
scale values can be analytically solved for in terms of a single 
overall mass scale $M$.  The expected size for $M$ depends on whether 
one adopts the naturalness or landscape pictures, allowing for the 
possibility of distinguishing between these two cases.  We also 
present possible variations of the model, and discuss more general 
implications of the landscape picture in this context.}

\end{center}
\end{titlepage}

\section{Introduction}
\label{sec:intro}

Weak scale supersymmetry is an extremely attractive idea.  It is 
based on a beautiful theoretical construction of enlarging the 
spacetime structure to anticommuting variables, and is supported 
indirectly by the successful unification of gauge couplings at high 
energies~\cite{Dimopoulos:1981zb}.  It also stabilizes the large 
hierarchy between the weak and the Planck scales due to a cancellation 
between the standard model and its superpartner contributions to 
the Higgs potential.  In fact, this latter property has been one 
of the strongest motivations for weak scale supersymmetry.

\ From the experimental point of view, the most exciting aspect of 
weak scale supersymmetry is the existence of various superpartners 
at the TeV scale.  Can we predict the spectrum of these superparticles? 
We already know, from the absence of a large new contribution to 
flavor changing neutral current and $CP$-violating processes, that 
the superparticle spectrum must have a certain special structure, 
such as flavor universality.  Moreover, non-discovery of both 
superparticles and a light Higgs boson at LEP~II puts strong constraints 
on the spectrum.  This typically leads to fine-tuning of order a few 
percent in reproducing the correct scale for electroweak symmetry 
breaking, and is called the supersymmetric fine-tuning problem (for 
a recent analysis, see~\cite{Kitano:2006gv}).  It seems plausible 
that successfully addressing this problem provides a key to the 
correct theory at the TeV scale, and to a fundamental mechanism 
or principle behind it.

There are two different approaches towards the supersymmetric 
fine-tuning problem.  A conventional approach is to search for 
a model that is ``natural.''  In the context of the minimal 
supersymmetric standard model (MSSM), this amounts to looking for 
a model in which the supersymmetry breaking mass-squared parameter 
for the up-type Higgs field, $m_{H_u}^2$, is somehow suppressed 
at the weak scale, since the electroweak scale is determined 
approximately by
\begin{equation}
  \frac{M_{\rm Higgs}^2}{2} = -m_{H_u}^2 - |\mu|^2,
\label{eq:ewsb}
\end{equation}
so that smaller $|m_{H_u}^2|$ requires a smaller amount of cancellation 
between the $m_{H_u}^2$ and $|\mu|^2$ terms, where $M_{\rm Higgs}$ and 
$\mu$ represent the physical Higgs boson mass and the supersymmetric 
Higgs-mass parameter, respectively.  On the other hand, there are lower 
bounds on the masses of superparticles, coming from the experimental 
bounds on the superparticle and the Higgs boson masses.  This leads to 
a nontrivial tension between the values of $m_{H_u}^2$ and other generic 
supersymmetry breaking squared masses $\tilde{m}^2$ --- typically 
it requires a small hierarchy between $|m_{H_u}^2|$ and $\tilde{m}^2$. 
In the context of gravity mediation -- arguably the ``simplest'' 
mediation of supersymmetry breaking -- this implies that we must find 
a model in which the ``tree-level'' and ``radiative'' contributions 
to $m_{H_u}^2$ cancel to a large extent, either ``accidentally,'' 
as in the scenario of~\cite{Feng:1999mn}, or by some mechanism, 
as in the model of~\cite{Choi:2005hd,Kitano:2005wc}.

An alternative approach towards the problem appears if we live in 
``the multiverse,'' rather than the universe.  Motivated partly by 
Weinberg's successful ``prediction'' of the observed value of the 
cosmological constant~\cite{Weinberg:1987dv}, and partly by the 
suggestion that string theory has an exponentially large number 
of discrete nonsupersymmetric vacua~\cite{Bousso:2000xa}, it has 
become increasingly plausible that our universe is only one among 
a tremendous number of various universes, in which physical 
constants can take vastly different values.  This ``landscape'' 
hypothesis may lead to a significant change in our notion of 
naturalness, and it is reasonable to consider the supersymmetric 
fine-tuning problem in this context.  It has recently been argued 
that the landscape picture may lead to a small hierarchy between 
the Higgs mass-squared parameter and the scale of superparticle masses 
$\tilde{m}$ under certain assumptions on the probability distributions 
of various couplings and $\tilde{m}$~\cite{Giudice:2006sn}. 
Specifically, under the existence of statistical ``pressures'' 
pushing $\tilde{m}$ towards larger values, the relation $v^2 
\sim \tilde{m}^2/8\pi^2$ may be obtained from environmental 
selection, where $v$ is the electroweak scale.%
\footnote{This conclusion depends on the probability distributions 
of parameters.  For example, if certain couplings do not ``scan,'' the 
low-energy theory may be split supersymmetry~\cite{Arkani-Hamed:2004fb}, 
or simply the standard model~\cite{Feldstein:2006ce}.  The assumption 
here corresponds to an independent scanning of $\tilde{m}$ and the 
supersymmetric couplings.  It is interesting that supersymmetry 
may still play an important role in addressing the gauge hierarchy 
problem even in the existence of a landscape of vacua, under 
certain mild assumptions.}
Moreover, if the parameter $\mu$ also scans independently with 
$\tilde{m}$ and if the holomorphic supersymmetry breaking Higgs 
mass-squared parameter, $\mu B$, is sufficiently small at a high 
scale, then we obtain $v^2 \sim |\mu|^2 \sim |m_{H_u}^2| \sim 
\tilde{m}^2/8\pi^2$.

It is interesting that the two different pictures described above 
can both lead to a scenario in which the supersymmetry breaking 
parameter $m_{H_u}^2$ crosses zero at a scale not much different from 
the weak scale.  In fact, the two pictures may not be totally unrelated. 
Suppose, for example, that the ultraviolet theory at the gravitational 
or unification scale gives universal scalar squared masses $m_0^2$ 
($>0$), as in the minimal supergravity scenario~\cite{Chamseddine:1982jx}. 
In this case, the parameter $m_{H_u}^2$ crosses zero at a renormalization 
scale of order the weak scale, as long as the gaugino masses are 
small compared with $|m_0^2|^{1/2}$.  This phenomenon is known 
as focus point behavior, and this class of theories was claimed to 
be natural~\cite{Feng:1999mn}, since $|m_{H_u}^2|$ is relatively small 
at the weak scale and thus no strong cancellation is required between 
the two terms in the right-hand-side of Eq.~(\ref{eq:ewsb}).  An immediate 
criticism of this argument, based on the conventional viewpoint, is 
that if the value of the top Yukawa coupling, $y_t$, were different, 
then the property of $|m_{H_u}^2|$ being small at the weak scale 
would be destroyed --- in other words, the fractional sensitivity 
of the weak scale, $v$, to a variation of the top Yukawa coupling, 
$\partial \ln v^2/\partial \ln y_t$, is very large.  This criticism, 
however, is not appropriate if the property of $|m_{H_u}^2| \ll 
\tilde{m}^2$ at the weak scale is a result of environmental selection. 
In this case, if $y_t$ were changed, the scale of supersymmetry 
breaking masses, $\tilde{m}$, would also be changed in such a way 
that $|m_{H_u}^2| \sim \tilde{m}^2/8\pi^2 \ll \tilde{m}^2$ {\it at 
the ``new'' weak scale} $\sim |m_{H_u}|$.  As a result, we always find 
$|m_{H_u}^2| \ll \tilde{m}^2$ at the ``weak scale'' {\it regardless 
of the value of $y_t$}.  The observed value of $y_t$ will then be 
determined as a result of (another) environmental selection, presumably 
a combination of the consideration in~\cite{Agrawal:1997gf} and others.

\ From the point of view of model-building, i.e. searching for the 
model describing physics above the TeV scale, we may then be motivated 
to look for a model in which $|m_{H_u}^2|$ is suppressed compared with 
$\tilde{m}^2$ at the weak scale, i.e. $|m_{H_u}^2|$ crosses zero at 
a scale close to the weak scale.  If this property arises without 
a strong cancellation between the ``tree-level'' and ``radiative'' 
contributions to $|m_{H_u}^2|$, then we can consider that the model is 
natural in the conventional sense.  Even if it arises due to a strong 
cancellation, however, the model may still be interesting since it can 
arise as a result of environmental selection under certain circumstances. 
Note that the requirement of $|m_{H_u}^2|$ being suppressed at the weak 
scale is different from the one that the Higgs mass-squared parameter, 
$|m_h^2| \simeq |m_{H_u}^2 + |\mu|^2|$, is suppressed at the weak scale, 
which should always be the case.  We are requiring that the cancellation 
(if any) must take place ``inside'' $m_{H_u}^2$, and not between 
$m_{H_u}^2$ and $|\mu|^2$.

Since the condition of $|m_{H_u}^2| \ll \tilde{m}^2$ at the weak scale 
gives only one constraint on the large number of soft supersymmetry 
breaking masses, we clearly need other guiding principles to narrow 
down the possibilities and obtain predictions on the superparticle masses. 
Without having a detailed knowledge of physics at the gravitational 
or unification scale, we simply take the viewpoint that the physics at 
that scale should be ``simple'' -- sufficiently simple that the resulting 
supersymmetry breaking masses also take a simple form.  This clearly 
makes sense if we take the conventional ``universe'' picture, and may 
also be supported by the absence of large supersymmetric flavor-changing 
and $CP$-violating contributions (which would arise if the superparticle 
masses were chaotic).  In the context of the ``multiverse'' (or landscape) 
picture, we merely hope that such a ``simple'' model is statistically 
preferred by the vacuum counting in the fundamental theory.  In practice, 
if a sufficiently ``simple'' model defined at the high energy scale 
gives $|m_{H_u}^2| \ll \tilde{m}^2$ at the weak scale, we consider 
it interesting regardless of the level of cancellation occurring in 
$m_{H_u}^2$.

In this paper we present an example of such models.  The model is very 
simple, and arises as a low-energy effective theory of higher dimensional 
theories in which the standard model gauge and Higgs fields propagate in 
the bulk while matter fields are confined on a $(3+1)$-dimensional brane. 
The compactification scale is of the order of the unification scale, 
and the low-energy effective theory below this scale is simply the 
MSSM.  Upon stabilizing a volume modulus by a simple gaugino condensation 
superpotential, the superparticle masses in the low-energy theory receive 
contributions from both moduli and anomaly mediations.  We find that 
this model gives vanishing $m_{H_u}^2$ at a scale (very) close to 
the weak scale, satisfying the criterion described above.  All the 
supersymmetry breaking parameters, except for the holomorphic Higgs 
mass-squared parameter, are predicted (essentially) in terms of a single 
overall mass parameter $M$, with the resulting spectrum showing a pattern 
distinct from conventional supergravity and gauge mediation models. 
This model gives a ``non-hierarchical'' spectrum of $M_\lambda \sim 
m_{\tilde{f}}$ ($= O(M)$), where $M_\lambda$ and $m_{\tilde{f}}$ represent 
generic gaugino and sfermion masses, although variations of the model 
giving the ``hierarchical'' spectrum of $M_\lambda \sim m_{\tilde{f}}/4\pi$ 
($\sim |\mu|$) may also be considered.  The scale of the overall mass 
parameter $M$ depends on which of the naturalness or landscape pictures 
we take, but will be generally in the range between $O(v)$ and a multi-TeV 
scale.  For the Higgs sector, we simply assume that the required structures 
for the $\mu$ and $\mu B$ parameters are prepared, presumably by statistical 
preference in the case that the landscape picture is adopted.

The paper is organized as follows.  In the next section we present 
our model and derive predictions on the supersymmetry breaking masses 
which are independent of the picture adopted.  In section~\ref{sec:impl} 
we discuss the implications of the model in both the ``universe'' 
and ``multiverse'' pictures, and argue that the difference can appear 
in the size of the overall mass scale for the superparticle masses. 
In section~\ref{sec:concl} we conclude by giving discussions on the issue 
of obtaining predictions for the superparticle masses in the landscape 
picture.  In particular, we present several possible scenarios arising 
from a landscape of vacua in the ``vicinity'' of the particular model 
in section~\ref{sec:model}, and elucidate under what conditions, or 
with what additional assumptions, the setup can give strong predictions 
on the superparticle spectrum.

\section{Model}
\label{sec:model}

In this section we present a simple model that has the property that 
the soft Higgs mass squared is vanishing at a scale close to the weak 
scale.  We consider that physics above the unification scale is higher 
dimensional, and that the standard model gauge and Higgs fields propagate 
in the bulk while the matter fields are localized on a $(3+1)$-dimensional 
brane.  The low-energy effective theory is then given by the following 
4D supergravity action:
\begin{eqnarray}
  S &=& \int\! d^4 x\, \sqrt{-g}\, \Biggl[ \int\! d^4\theta\, 
    C^\dagger C \Bigl( -3 (T+T^\dagger) 
      + (T+T^\dagger) H^\dagger H + M^\dagger M \Bigr)
\nonumber\\
  && {} + \Biggl\{ \int\! d^2\theta\, \Bigl( \frac{1}{4} T 
    {\cal W}^{a \alpha} {\cal W}^a_\alpha + C^3 W \Bigr) + {\rm h.c.}
    \Biggr\} \Biggl],
\label{eq:action}
\end{eqnarray}
where $C$ is the chiral compensator superfield, $T$ is the moduli 
superfield parameterizing the volume of the compact dimensions, and 
$g_{\mu\nu}$ is the metric in the superconformal frame.  The superfields 
$H$ and $M$ collectively represent the Higgs and matter fields of the 
MSSM, i.e. $H = H_u, H_d$ and $M = Q_i, U_i, D_i, L_i, E_i$ with $i$ 
the generation index, and the superpotential $W$ contains the usual 
MSSM Yukawa couplings $W_{\rm Yukawa}$.  This setup naturally arises, 
for example, if grand unification is realized in higher dimensions 
above the compactification scale~\cite{Kawamura:2000ev}.%
\footnote{In the case of 5D $SU(5)$ with matter localized on the $SU(5)$ 
brane, the volume of the compact extra dimension cannot be much larger 
than the cutoff scale to avoid excessive proton decay caused by the 
exchange of the unified gauge bosons.  Alternatively, the matter 
fields can be located in the bulk, with the zero-mode wavefunctions 
localized strongly towards the $SU(5)$-violating brane.  This reproduces 
the action of Eq.~(\ref{eq:action}) at low energies while preserving 
the $SU(5)$ understanding of the matter quantum numbers.}
In Eq.~(\ref{eq:action}), we have assumed that moduli fields other 
than $T$, e.g. ones parameterizing the shape of the compact dimensions, 
(if any) are absent in the low-energy theory.  We have also assumed 
that higher order terms, e.g. terms involving powers of $1/(T+T^\dagger)$, 
are sufficiently suppressed, which is technically natural since the 
theory is weakly coupled at the compactification scale.

To obtain realistic phenomenology at low energies, the moduli field 
$T$ must be stabilized.  We assume that the stabilization superpotential 
for $T$ takes the simple form arising from a single gaugino condensation. 
The superpotential $W$ is then given by
\begin{equation}
  W = W_{\rm Yukawa} + A e^{-a T} + c,
\label{eq:W}
\end{equation}
where $a$ is a real constant.  The parameters $A$ and $c$ are constants 
of order unity and the gravitino mass ($\ll 1$), respectively (in units 
of the 4D gravitational constant $M_{\rm Pl} \simeq 10^{18}~{\rm GeV}$, 
which is taken to be $1$).  These parameters can be taken real in the 
presence of an approximate shift symmetry for ${\rm Im} T$.  Since 
the superpotential of Eq.~(\ref{eq:W}) stabilizes the modulus $T$ 
at a supersymmetry preserving anti-de~Sitter vacuum, with $\langle 
T+T^\dagger \rangle \simeq 2 a^{-1}\ln(a/c)$, we need an uplifting 
(supersymmetry breaking) potential, which we take to be independent 
of $T$ in the superconformal basis:
\begin{equation}
  \delta S = - \int\! d^4 x\, \sqrt{-g}\, \int\! d^4\theta\, 
    C^{\dagger 2} C^2 \theta^2 \bar{\theta}^2 d,
\label{eq:uplift}
\end{equation}
where $d$ is a positive constant.  A term of this form effectively 
arises from almost any supersymmetry breaking occurring in the 
$(3+1)$-dimensional subspace, which we assume to be sequestered 
from the observable sector.  (The case without sequestering will be 
discussed in section~\ref{sec:concl}.)  In fact, this setup can arise 
as the low-energy effective theory of the string theory scenario 
discussed in Ref.~\cite{Kachru:2003aw}.  In that context, the constant 
$c$ arises from fluxes stabilizing the moduli other than $T$, and 
$d$ from the vacuum energy associated with $\overline{D3}$ branes, 
located at the bottom of a warped throat.  (The configuration of 
the gauge, Higgs and matter fields described before corresponds to 
identifying them as $D7$-, $D7$- and $D3$-brane fields, respectively.)

The minimization of the potential, derived from Eqs.~(\ref{eq:action}~%
--~\ref{eq:uplift}), leads to supersymmetry breaking ($F$-term) 
expectation values for the compensator $C$ and the modulus $T$:
\begin{eqnarray}
  \frac{F_C}{C} &=& \frac{c}{(T+T^\dagger)^{3/2}} 
  \,\,=\,\, m_{3/2},
\label{eq:F_C} \\
  \frac{F_T}{T+T^\dagger} &=& \frac{2}{a(T+T^\dagger)} m_{3/2} 
  \,\,\equiv\,\, M_0,
\label{eq:F_T}
\end{eqnarray}
where $m_{3/2}$ is the gravitino mass.  This implies that there is 
a little hierarchy between the sizes of $F_C$ and $F_T$:
\begin{equation}
  \frac{F_C/C}{F_T/(T+T^\dagger)} 
  \,=\, \frac{a}{2}(T+T^\dagger) 
  \,=\, \ln\biggl(\frac{M_{\rm Pl}}{m_{3/2}}\biggr),
\label{eq:LH}
\end{equation}
so that the supersymmetry breaking parameters in the MSSM 
receive comparable contributions from both moduli and anomaly 
mediations~\cite{Choi:2004sx}.  Here, we have recovered the 
gravitational constant $M_{\rm Pl}$ in the right-hand-side of 
Eq.~(\ref{eq:LH}).  Note that the above Eqs.~(\ref{eq:F_C}~%
--~\ref{eq:LH}) are valid up to corrections of $O(1/8\pi^2) = 
O(1/\ln(M_{\rm Pl}/m_{3/2}))$.

The supersymmetry breaking masses in the present model show the 
behavior of a reduced effective messenger scale, $M_{\rm mess}$, 
due to an interplay between the moduli and anomaly mediated 
contributions~\cite{Choi:2005uz} (for a simple proof, 
see~\cite{Kitano:2006gv}).  By solving renormalization group 
equations at the one-loop level, the soft supersymmetry breaking 
masses at an arbitrary renormalization scale $\mu_R$ are given by
\begin{eqnarray}
  M_a(\mu_R) &=& M_0 \Biggl[ 1 - \frac{b_a}{8\pi^2} g_a^2(\mu_R) 
    \ln\Biggl( \frac{M_{\rm mess}}{\mu_R} \Biggr) \Biggr],
\label{eq:Ma-LE} \\
  m_I^2(\mu_R) &=& M_0^2 \Biggl[ r_I - 4\, \Biggl\{ 
    \gamma_I(\mu_R) - \frac{1}{2} \frac{d\gamma_I(\mu_R)}{d \ln\mu_R} 
    \ln\Biggl( \frac{M_{\rm mess}}{\mu_R} \Biggr) \Biggr\} 
    \ln\Biggl( \frac{M_{\rm mess}}{\mu_R} \Biggr) \Biggr],
\label{eq:mi2-LE} \\
  A_{IJK}(\mu_R) &=& M_0 \Biggl[ -(r_I+r_J+r_K) 
    + 2 \Bigl\{ \gamma_I(\mu_R)+\gamma_J(\mu_R)+\gamma_K(\mu_R) \Bigr\} 
    \ln\Biggl( \frac{M_{\rm mess}}{\mu_R} \Biggr) \Biggr],
\label{eq:Aijk-LE}
\end{eqnarray}
where $M_a$, $m_I^2$ and $A_{IJK}$ are gaugino masses, non-holomorphic 
scalar squared masses, and scalar trilinear interactions (with the Yukawa 
couplings factored out), respectively.  The indices $I,J,K$ run over 
$Q_i$, $U_i$, $D_i$, $L_i$, $E_i$, $H_u$, $H_d$, with $r_I$'s defined 
by $r_{Q_i} = r_{U_i} = r_{D_i} = r_{L_i} = r_{E_i} = 0$ and $r_{H_u} 
= r_{H_d} = 1$;%
\footnote{The notation here follows that of Ref.~\cite{Kitano:2005wc} 
except that the sign convention for $A_{IJK}$ is reversed.}
$g_a(\mu_R)$ are the running gauge couplings at a scale $\mu_R$, and 
$b_a$ and $\gamma_I(\mu_R)$ are the beta-function coefficients and the 
anomalous dimensions, respectively, defined by $d(1/g_a^2)/d\ln\mu_R 
= -b_a/8\pi^2$ and $d \ln Z_I/d \ln\mu_R = -2 \gamma_I$, where $Z_I$ 
is the wavefunction renormalization factor for the field $I$.  The 
parameter $M_{\rm mess}$ is given by
\begin{equation}
  M_{\rm mess} = f \frac{M_U}{(M_{\rm Pl}/m_{3/2})^{1/2}},
\label{eq:eff-mess}
\end{equation}
where $M_U$ represents the compactification scale, which is of the order 
of the unification scale $\approx 10^{16}~{\rm GeV}$, and $f$ is an $O(1)$ 
coefficient depending, e.g., on $A$ in Eq.~(\ref{eq:W}).  The parameter 
$M_0$ is defined in Eq.~(\ref{eq:F_T}) and represents the overall mass 
scale for the supersymmetry breaking parameters.

The expressions of Eqs.~(\ref{eq:Ma-LE}~--~\ref{eq:Aijk-LE}) show that 
the supersymmetry breaking masses in this model take a very simple form:
\begin{equation}
  M_1 = M_2 = M_3 = M_0,
\label{eq:gaugino-masses}
\end{equation}
\begin{equation}
  m_{\tilde{Q}_i}^2 = m_{\tilde{U}_i}^2 = m_{\tilde{D}_i}^2 
    = m_{\tilde{L}_i}^2 = m_{\tilde{E}_i}^2 = 0,
\qquad
  m_{H_u}^2 = m_{H_d}^2 = M_0^2,
\label{eq:scalar-masses}
\end{equation}
\begin{equation}
  A_u = A_d = A_e = -M_0,
\label{eq:Aterms}
\end{equation}
at the effective messenger scale
\begin{equation}
  M_{\rm mess} \simeq \sqrt{M_U M_0} = O(10^9\!\sim\!10^{10}~{\rm GeV}),
\label{eq:M_mess}
\end{equation}
where we have denoted the squark and slepton squared masses as 
$m_{\tilde{F}}^2$ ($F=Q_i,U_i,D_i,L_i,E_i$) and the scalar trilinear 
interaction parameters, which are flavor universal in the present 
model, as $A_u$, $A_d$ and $A_e$.  (Our sign convention for the 
soft supersymmetry breaking parameters follows that of the SUSY 
Les Houches Accord~\cite{Skands:2003cj}.)  Here, we have suppressed 
possible higher order corrections of $O(M_0^2/8\pi^2)$ in 
Eq.~(\ref{eq:scalar-masses}).%
\footnote{Approximate flavor universality for these corrections must 
be assumed in the case that $M_0$ is not much larger than a TeV.}
Note that the spectrum of Eqs.~(\ref{eq:gaugino-masses}~--~\ref{eq:Aterms}) 
is identical with what would be obtained at the compactification scale 
in simple moduli mediated (or equivalently Scherk-Schwarz) supersymmetry 
breaking~\cite{Barbieri:2001yz}.  The low-energy soft supersymmetry 
breaking parameters, defined at the weak scale $m_{\rm w}$, are then 
given by evolving Eqs.~(\ref{eq:gaugino-masses}~--~\ref{eq:Aterms}) 
down from $M_{\rm mess}$ to $m_{\rm w}$, or simply by using 
Eqs.~(\ref{eq:Ma-LE}~--~\ref{eq:mi2-LE}) for $\mu_R = m_{\rm w}$.

\begin{figure}[t]
\begin{center}
  \includegraphics[height=7.5cm]{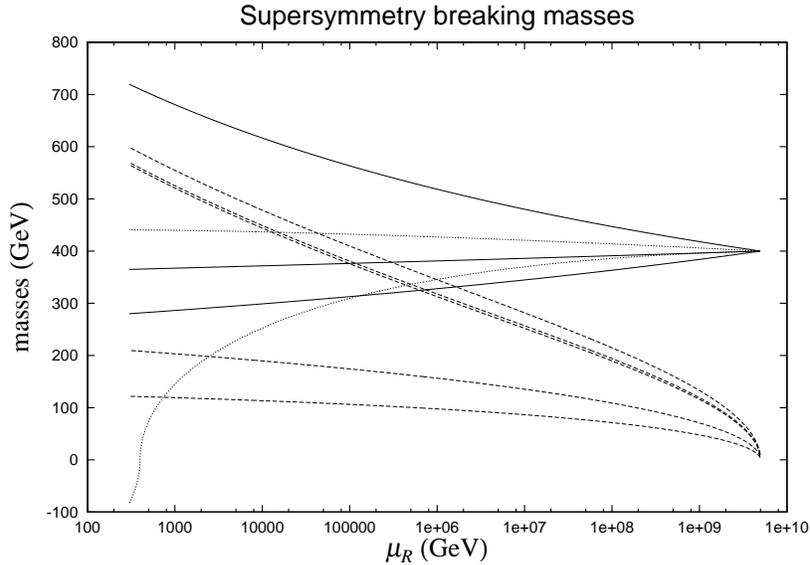}
\end{center}
\caption{Evolutions of soft supersymmetry breaking masses below 
 $M_{\rm mess} = 5 \times 10^9~{\rm GeV}$ for $M_0 = 400~{\rm GeV}$ 
 and $\tan\beta = 10$.  Solid lines represent the gaugino masses 
 ($M_3$, $M_2$ and $M_1$ from the top), dashed lines the first 
 two generation sfermion masses ($m_{\tilde{Q}}$, $m_{\tilde{U}}$, 
 $m_{\tilde{D}}$, $m_{\tilde{L}}$ and $m_{\tilde{E}}$ from the 
 top), and dotted lines the Higgs mass parameter ($m_{H_d}$ and 
 $m_{H_u}$ from the top).  Here, $m_\Phi$ ($\Phi = \tilde{Q}, 
 \tilde{U}, \tilde{D}, \tilde{L}, \tilde{E}, H_u, H_d$) is defined 
 by $m_\Phi \equiv {\rm sgn}(m_\Phi^2) |m_\Phi^2|^{1/2}$.  The 
 pole mass for the top quark is chosen to be the central value 
 of the recently reported range $m_t = 171.4 \pm 2.1~{\rm 
 GeV}$~\cite{Brubaker:2006xn}.}
\label{fig:model}
\end{figure}
In Fig.~\ref{fig:model}, we show the evolutions of the soft supersymmetry 
breaking parameters in the present model, taking $M_0 = 400~{\rm GeV}$, 
$M_{\rm mess} = 5 \times 10^9~{\rm GeV}$ and $\tan\beta \equiv \langle 
H_u \rangle/\langle H_d \rangle = 10$ for illustrative purposes. 
In the figure, we have taken the supersymmetry breaking masses 
of Eqs.~(\ref{eq:gaugino-masses}~--~\ref{eq:Aterms}) at the scale 
$M_{\rm mess}$, and evolved them down using the one-loop renormalization 
group equations of the MSSM.  (The two-loop renormalization group 
equations have been used for the supersymmetric parameters.)  Note 
that while the soft supersymmetry breaking parameters are depicted 
only for $\mu_R \leq M_{\rm mess}$, it should be understood that they 
are, in fact, generated at a scale of order $M_U$.  (The squark and 
slepton squared masses are negative at scales above $M_{\rm mess}$, 
but this does not cause a problem since our vacuum is metastable at the 
time scale of the age of the universe.)  Below, we will choose $M_0$ 
and $M_{\rm mess}$ to be free parameters of our analysis, since these 
parameters have $O(1)$ uncertainties that cannot be determined from 
the low-energy data alone.  The value of $\tan\beta$ is determined 
by the Higgs sector parameters, $\mu$ and $\mu B$, whose origin 
we leave unspecified.%
\footnote{We note that essentially all the conclusions below also apply 
in any theory in which the soft supersymmetry breaking masses take the 
form of Eqs.~(\ref{eq:gaugino-masses}~--~\ref{eq:Aterms}) at the scale 
of Eq.~(\ref{eq:M_mess}).  These boundary conditions might arise, e.g., 
in a theory where the fundamental scale is at an intermediate scale 
or in a theory where there is a physical threshold at an intermediate 
scale.}

A remarkable feature of the superparticle masses in Fig.~\ref{fig:model} 
is that the up-type Higgs mass-squared parameter crosses zero at the 
superparticle mass scale:
\begin{equation}
  m_{H_u}^2(\mu_C) = 0
  \qquad {\rm at} \qquad \mu_C \simeq M_0.
\label{eq:mHu2}
\end{equation}
While the precise value of $\mu_C$ -- the scale where $m_{H_u}^2$ crosses 
zero -- depends on the values of $M_{\rm mess}$ and $\tan\beta$, it is 
of order $M_0$ for a wide range of these parameters.  Note that $\mu_C$ 
does not depend on $M_0$, since the renormalization group equations are 
homogeneous in $M_0$.  (If we take $M_U$ to be a free parameter, instead 
of $M_{\rm mess}$, then $\mu_C$ depends slightly on $M_0$ for a fixed 
$M_U$, through a weak dependence of $M_{\rm mess}$ on $M_0$.)  In the 
example of $M_0 = 400~{\rm GeV}$ in Fig.~\ref{fig:model}, the value of 
$\mu_C$ is within a factor of $2$ from $M_0$ for $M_{\rm mess} \approx 
(10^9\!\sim\!10^{10})~{\rm GeV}$ for $\tan\beta \approx (5\!\sim\!30)$. 
(In fact, a value of $M_{\rm mess}$ giving $\mu_C$ within a factor 
of $2$ from $M_0$ can be found for $\tan\beta \approx (3\!\sim\!50)$. 
The mass squared for the right-handed stau, however, becomes negative 
at the weak scale for $\tan\beta \simgt 30$.)  These results do not 
change significantly by including higher order effects, 
e.g. the two-loop renormalization group effects, or by varying the 
top quark mass within a 2$\sigma$ range of the recently reported 
value, $m_t = 171.4 \pm 2.1~{\rm GeV}$~\cite{Brubaker:2006xn}. 
At the leading order, we find from Eq.~(\ref{eq:mi2-LE}) that 
the scale $\mu_C$ is given by
\begin{eqnarray}
  \mu_C &\approx& M_{\rm mess} 
    \exp\Biggl( \frac{8\pi^2\Bigl( 6 y_t^2 - 3 g_2^2 
      - \sqrt{64 g_3^2 y_t^2 - 36 y_t^4 + 15 g_2^4} \Bigr)}
      {32 g_3^2 y_t^2 - 36 y_t^4 + 18 g_2^2 y_t^2 + 3 g_2^4} \Biggr)
\nonumber\\
  &\approx& 10^{-7} M_{\rm mess},
\label{eq:mu_C}
\end{eqnarray}
where the top Yukawa coupling, $y_t$, and the $SU(3)_C$ and $SU(2)_L$ 
gauge couplings, $g_3$ and $g_2$, are evaluated at the scale $\mu_R 
\simeq \mu_C$, and we have neglected the small effects from the bottom 
Yukawa coupling, $y_b$, and the $U(1)_Y$ gauge coupling, $g_1$.  To 
obtain $\mu_C \simeq M_0$, a larger $M_0$ requires a larger $M_{\rm mess} 
\propto M_0$.  For fundamental parameters of the theory, this implies 
$f \propto M_0$ (see Eq.~(\ref{eq:eff-mess})).

Since the superparticle mass scale $M_0$ is close to $\mu_C$, we can 
evaluate the soft supersymmetry breaking parameters at the superparticle 
mass scale $M_0$ approximately by substituting Eq.~(\ref{eq:mu_C}) 
into Eqs.~(\ref{eq:gaugino-masses}~--~\ref{eq:Aterms}).  This gives 
predictions for {\it all} the supersymmetry breaking masses, except for 
the holomorphic Higgs mass-squared parameter $\mu B$ (and $m_{H_u}^2$), 
in terms of the overall mass scale $M_0$ and the running gauge and Yukawa 
couplings at that scale.  Note that we even do not have to know the 
value of $M_{\rm mess}$ -- for given values of $M_0$ and $\tan\beta$, 
which we need to obtain the values of the Yukawa couplings, we can 
predict all the supersymmetry breaking parameters with the assumption 
of Eq.~(\ref{eq:mHu2}). 

\begin{figure}[t]
\begin{center}
  \includegraphics[height=5.8cm]{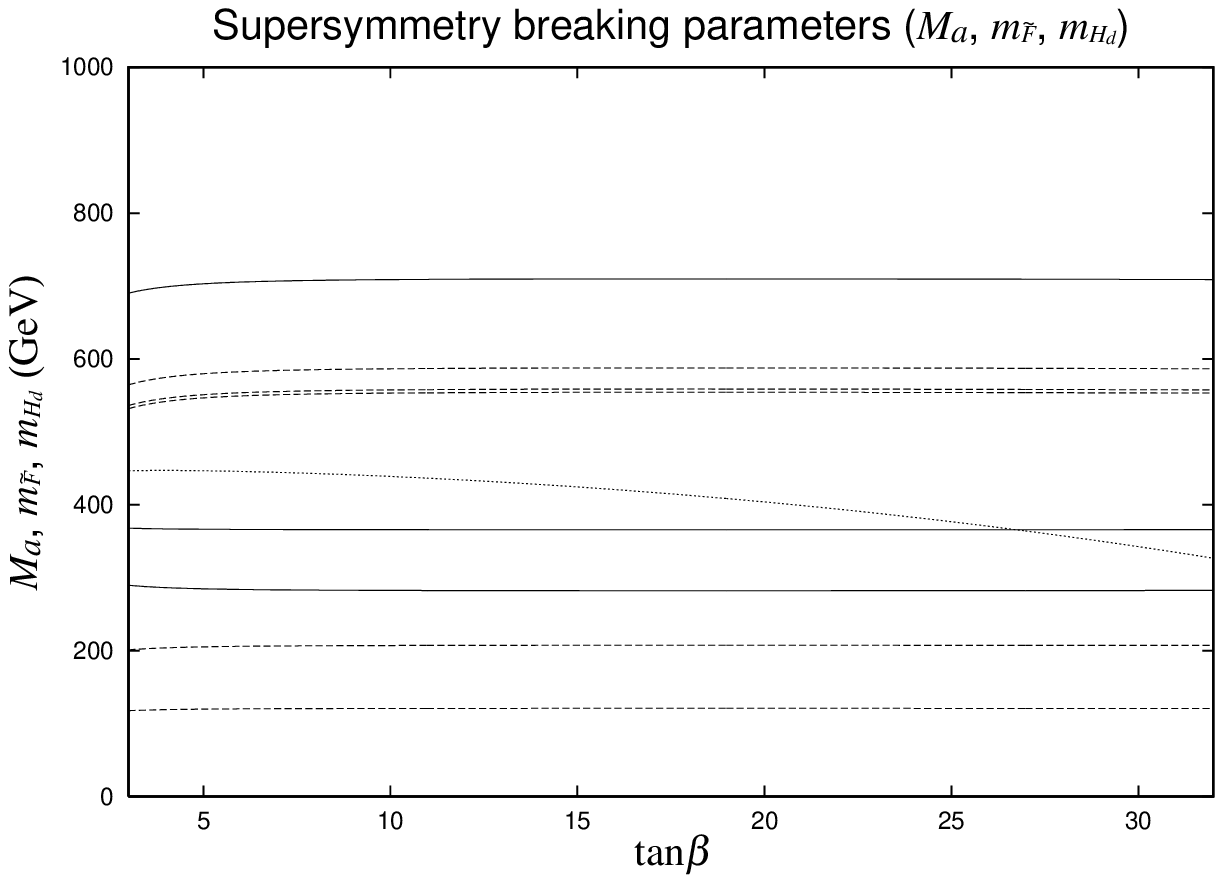}
\hspace{3mm}
  \includegraphics[height=5.8cm]{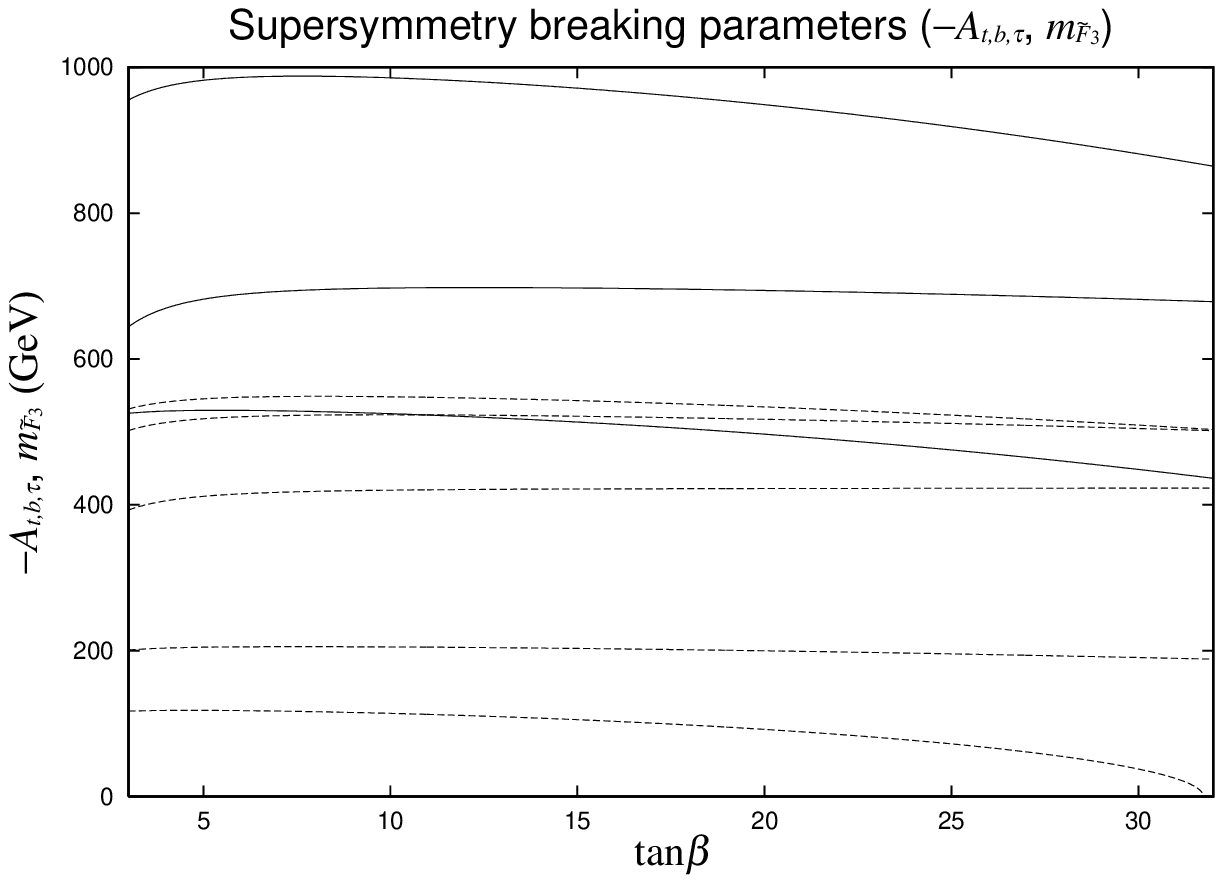}
\end{center}
\caption{Predictions for the soft supersymmetry breaking parameters 
 as a function of $\tan\beta$ for $M_0 = 400~{\rm GeV}$.  The left panel 
 shows the predictions for the gaugino masses (solid; $M_3$, $M_2$ and 
 $M_1$ from the top), the first two generation sfermion masses (dashed; 
 $m_{\tilde{Q}}$, $m_{\tilde{U}}$, $m_{\tilde{D}}$, $m_{\tilde{L}}$ and 
 $m_{\tilde{E}}$ from the top), and the down-type Higgs boson mass $m_{H_d}$ 
 (dotted).  The right panel shows those for the third generation scalar 
 trilinear interaction parameters (solid; $A_b$, $A_t$ and $A_\tau$ from 
 the top) and the third generation sfermion masses (dashed; $m_{\tilde{D}_3}$, 
 $m_{\tilde{Q}_3}$, $m_{\tilde{U}_3}$, $m_{\tilde{L}_3}$ and $m_{\tilde{E}_3}$ 
 from the top).  For $A_t$, $A_b$ and $A_\tau$, which are negative, the 
 absolute values are plotted.  The scalar trilinear interaction parameters 
 for the first two generations, $A_u$, $A_d$ and $A_e$, are not shown.}
\label{fig:masses}
\end{figure}

In Fig.~\ref{fig:masses}, we present the predicted values of the 
supersymmetry breaking parameters for $M_0 = 400~{\rm GeV}$ as 
a function of $\tan\beta$.
The left panel shows the predictions of the gaugino masses, $M_a$, 
the first two generation sfermion masses, $m_{\tilde{F}}$, and the 
down-type Higgs boson mass, $m_{H_d}$.  The right panel shows the third 
generation scalar trilinear interaction parameters, $A_{t,b,\tau}$, and 
the third generation sfermion masses, $m_{\tilde{F}_3}$.  The scalar 
trilinear interaction parameters for the first two generations, $A_{u,d,e}$, 
are not shown.  The predictions for $M_a$, $m_{\tilde{F}}$, $m_{\tilde{Q}_3}$, 
$m_{\tilde{U}_3}$, $m_{\tilde{L}_3}$, $A_t$ (and $A_u$, which is not shown) 
are rather insensitive to the value of $\tan\beta$, while those for 
$m_{H_d}$, $m_{\tilde{D}_3}$, $m_{\tilde{E}_3}$, $A_b$, $A_\tau$ (and 
$A_d$, $A_e$) have weak sensitivities to $\tan\beta$.  (The sensitivity 
is strong for $m_{\tilde{E}_3}$ for $\tan\beta \simgt 30$ where it 
approaches zero.)  For $\tan\beta \sim 10$, the predicted ratios among 
the soft supersymmetry breaking parameters (including the first two 
generation scalar trilinear interaction parameters) are given by
\begin{eqnarray}
  && M_1 : M_2 : M_3 
\quad:\quad 
  m_{\tilde{Q}} : m_{\tilde{U}} : m_{\tilde{D}} : 
    m_{\tilde{L}} : m_{\tilde{E}} 
\quad:\quad 
  m_{\tilde{Q}_3} : m_{\tilde{U}_3} : m_{\tilde{D}_3} : 
    m_{\tilde{L}_3} : m_{\tilde{E}_3} 
\quad: 
\nonumber\\
  && \qquad\qquad 
  m_{H_d} 
\quad:\quad 
  -A_u : -A_d : -A_e 
\quad:\quad 
  -A_t : -A_b : -A_\tau 
\nonumber\\
  && \,\,\simeq\,\, 0.71 : 0.91 : 1.8 
\quad:\quad 
  1.5 : 1.4 : 1.4 : 0.52 : 0.30 
\quad:\quad 
  1.3 : 1.1 : 1.4 : 0.51 : 0.28 
\quad:
\nonumber\\
  && \qquad\qquad 
  1.1 
\quad:\quad 
  2.2 : 2.6 : 1.3
\quad:\quad 
  1.7 : 2.5 : 1.3.
\label{eq:prediction}
\end{eqnarray}
Here, we have presented the numbers in units of $M_0$.  Note that these 
numbers are subject to errors of $O(10\%)$, coming from ``higher order'' 
effects, for quantities associated with the colored superparticles. 
(The errors for quantities that are not associated with the colored 
superparticles are smaller.)  In the case that we take the ``universe'' 
picture, these effects include the fact that the superparticle mass scale 
$M_0$ does not ``coincide'' with $\mu_C$, although the two are of the 
same order.  This source of errors does not exist if we adopt the 
``multiverse'' picture, where $M_0$ and $\mu_C$ are very close.

The predictions of Eq.~(\ref{eq:prediction}) have a sensitivity to 
the value of $M_0$, but only through the running of the gauge and 
Yukawa couplings.  As a result, these predictions, and the predictions 
for the ratios of the supersymmetry breaking masses obtained from 
Fig.~\ref{fig:masses}, are valid in a wide range of $M_0$ with only 
small corrections.  In the case that $M_0$ is in a multi-TeV region 
(as will be the case in the ``multiverse'' picture; see the next 
section), the corrections are still smaller than about $10\%$. 
For example, the predictions of Eq.~(\ref{eq:prediction}) change 
for $M_0 = 3~{\rm TeV}$ to
\begin{eqnarray}
  && M_1 : M_2 : M_3 
\quad:\quad 
  m_{\tilde{Q}} : m_{\tilde{U}} : m_{\tilde{D}} : 
    m_{\tilde{L}} : m_{\tilde{E}} 
\quad:\quad 
  m_{\tilde{Q}_3} : m_{\tilde{U}_3} : m_{\tilde{D}_3} : 
    m_{\tilde{L}_3} : m_{\tilde{E}_3} 
\quad: 
\nonumber\\
  && \qquad\qquad 
  m_{H_d} 
\quad:\quad 
  -A_u : -A_d : -A_e 
\quad:\quad 
  -A_t : -A_b : -A_\tau 
\nonumber\\
  && \,\,\simeq\,\, 0.63 : 0.90 : 1.8 
\quad:\quad 
  1.5 : 1.4 : 1.4 : 0.56 : 0.33 
\quad:\quad 
  1.3 : 1.0 : 1.4 : 0.56 : 0.31 
\quad:
\nonumber\\
  && \qquad\qquad 
  1.1 
\quad:\quad 
  2.2 : 2.7 : 1.4
\quad:\quad 
  1.8 : 2.5 : 1.4,
\label{eq:prediction-2}
\end{eqnarray}
but these are not much different from the ones in Eq.~(\ref{eq:prediction}). 

We finally discuss the Higgs sector of the model.  To have the correct 
electroweak symmetry breaking phenomenology, the $\mu$ and $\mu B$ 
parameters must be of order the weak scale.  In particular, the classical 
contribution to $B \equiv \mu B/\mu$ of order the gravitino mass must 
be suppressed.  Here we simply assume that the value of $B$ is sufficiently 
suppressed, for example the case that $B$ is somehow dominated by the 
quantum (anomalous) contribution: $B = 2 M_0 \{\gamma_{H_u}(\mu_R) + 
\gamma_{H_d}(\mu_R)\} \ln(M_{\rm mess}/\mu_R)$.  (This expression for $B$ 
is, in fact, a solution to the one-loop renormalization group equation.) 
We may also consider the case that $\mu$ is generated by the expectation 
value of a singlet field through $W = \lambda S H_u H_d$ (at least in 
the context of the ``universe'' picture), whose effect on the evolutions 
of the Higgs soft masses are suppressed if the value of $\lambda$ is 
sufficiently small.

\section{Implications}
\label{sec:impl}

We have seen that the model given by Eqs.~(\ref{eq:action},~\ref{eq:W},%
~\ref{eq:uplift}) provides the predictions of Eq.~(\ref{eq:prediction}), 
which depend only very weakly on the values of $\tan\beta$ and $M_0$ 
(see Fig.~\ref{fig:masses} and Eq.~(\ref{eq:prediction-2})).  The 
expected range for the overall scale $M_0$, however, differs depending 
on the scenario we consider. In this section we discuss this issue, 
as well as other phenomenological implications of the model.

\begin{figure}[t]
\begin{center}
  \includegraphics[height=5.8cm]{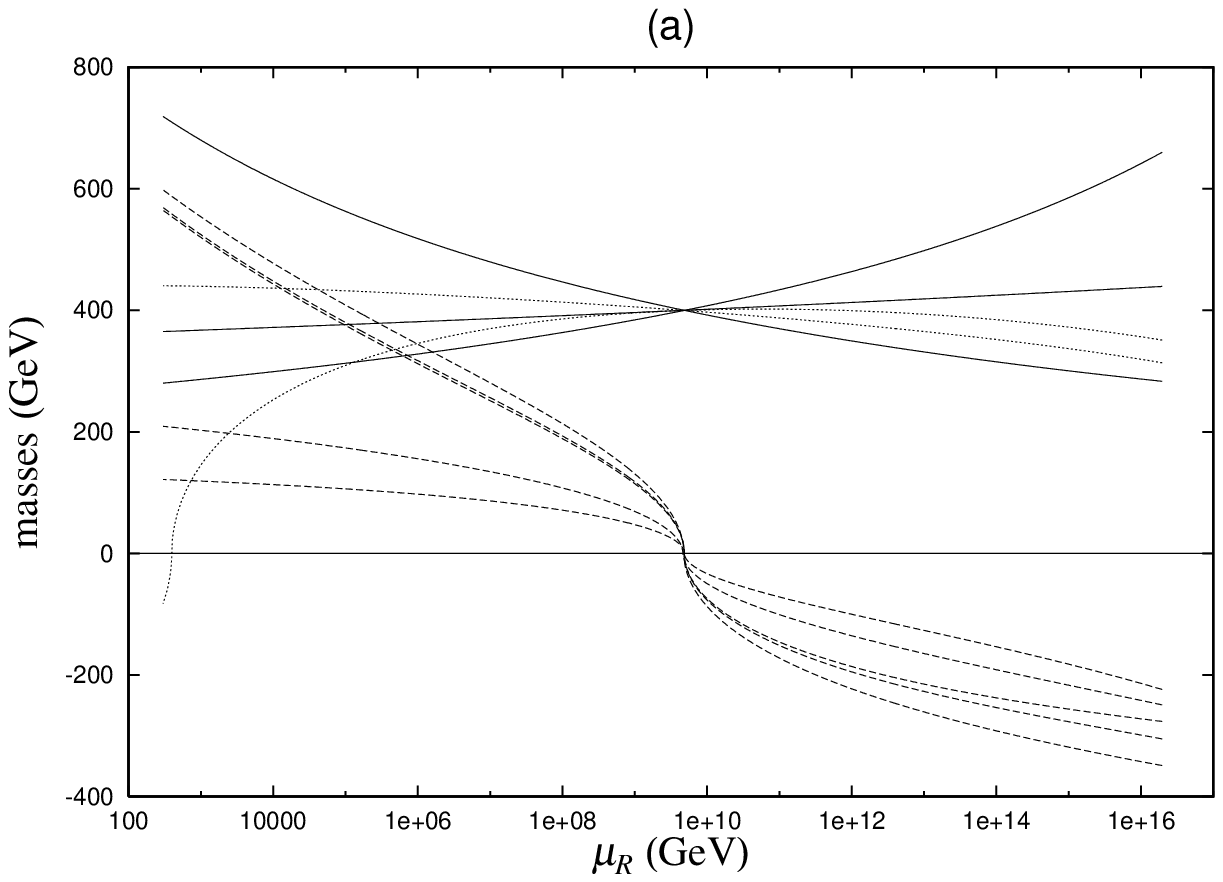}
\hspace{3mm}
  \includegraphics[height=5.8cm]{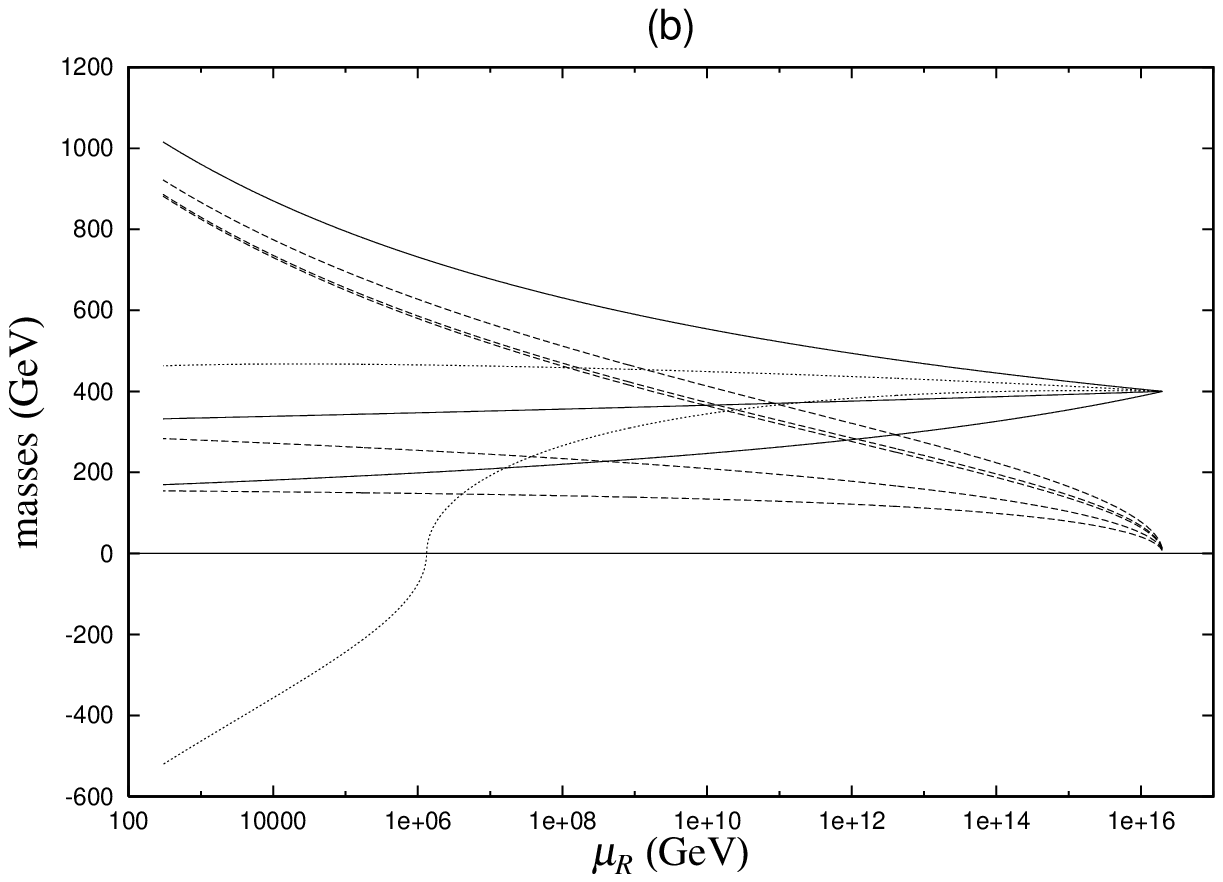}
\end{center}
\caption{Evolutions of soft supersymmetry breaking masses in 
 the model of section~\ref{sec:model} (the left panel), and 
 in the model where the supersymmetry breaking parameters of 
 Eqs.~(\ref{eq:gaugino-masses}~--~\ref{eq:Aterms}) are given 
 at the unification scale, $M_U \approx 10^{16}~{\rm GeV}$ (the 
 right panel).  Here, we have taken $M_0 = 400~{\rm GeV}$ and 
 $\tan\beta = 10$ in both cases.  Solid lines represent the gaugino 
 masses ($M_3$, $M_2$ and $M_1$ from the top), dashed lines the first 
 two generation sfermion masses ($m_{\tilde{Q}}$, $m_{\tilde{U}}$, 
 $m_{\tilde{D}}$, $m_{\tilde{L}}$ and $m_{\tilde{E}}$ from the 
 top), and dotted lines the Higgs mass parameter ($m_{H_d}$ and 
 $m_{H_u}$ from the top).}
\label{fig:comp}
\end{figure}

Let us first take the conventional ``universe'' picture, i.e. the 
overall scale $M_0$ does not effectively ``scan.''  In this case, 
our guiding principle will be ``naturalness,'' i.e. the observed 
scale of electroweak symmetry breaking, $v \simeq 174~{\rm GeV}$, 
should be a ``typical'' value in the parameter space of the model. 
For fixed values of the supersymmetric couplings, this is rather 
clear in our model because of the suppression of $m_{H_u}^2$ relative 
to the other soft masses at the weak scale.  (We assume that the Higgs 
sector is arranged such that there is no large $\mu B$ term of order 
$\mu m_{3/2}$.)  Naturalness of the model becomes clearer when compared 
with other, typical supersymmetry breaking models.  Consider, for 
example, a model in which the supersymmetry breaking parameters of 
Eqs.~(\ref{eq:gaugino-masses}~--~\ref{eq:Aterms}) are generated at 
the unification scale, $M_U \approx 10^{16}~{\rm GeV}$, as in the 
pure moduli mediated model of~\cite{Barbieri:2001yz}.  In this case, 
the size of the up-type Higgs mass squared $|m_{H_u}^2|$, relative 
to the other soft masses, is much larger at the weak scale. 
(The evolutions of soft masses in the two models are depicted 
in Fig.~\ref{fig:comp}.) 
An important point is that while $|m_{H_u}^2|$ keeps increasing 
towards the infrared from the scale $\mu_C$ where $m_{H_u}^2$ 
crosses zero, dragged by increasing $M_3$ through $g_3$ and $y_t$, 
the right-handed slepton masses $m_{\tilde{E}}$ stay almost constant, 
as they receive only small contributions through $g_1$.  As a 
consequence, if the crossing scale $\mu_C$ is much larger than the 
weak scale, we would obtain a hierarchy $|m_{H_u}^2|/m_{\tilde{E}}^2 
\gg 1$ at the weak scale (see Fig.~\ref{fig:comp}b), leading 
to fine-tuning between the $m_{H_u}^2$ and $|\mu|^2$ terms in 
Eq.~(\ref{eq:ewsb}) under the LEP~II constraint of $m_{\tilde{E}} 
\simgt 100~{\rm GeV}$.  Our model avoids this because $\mu_C$ is 
close to the weak scale (see Fig.~\ref{fig:comp}a).

\begin{figure}[t]
\begin{center}
  \includegraphics[height=7.5cm]{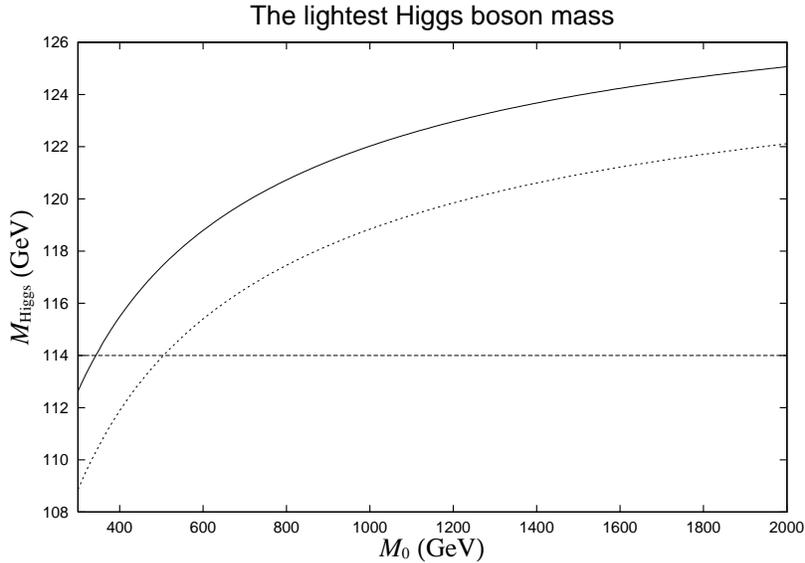}
\end{center}
\caption{Physical Higgs boson mass $M_{\rm Higgs}$ as a function of $M_0$ 
 for $\tan\beta = 5$ (dotted line) and $\tan\beta = 10$ (solid line). 
 The $\mu$ parameter is chosen to be $\mu = 150~{\rm GeV}$.  The horizontal 
 dashed line represents the experimental lower bound of $M_{\rm Higgs} 
 \simeq 114~{\rm GeV}$.}
\label{fig:Higgs}
\end{figure}

Since there is no particular reason that $\mu_C$ is extremely close 
to the scale of superparticle masses, $|\mu_C-M_0|/M_0 \ll 1$, we 
expect that there is some discrepancy between the two quantities, 
e.g. $|\ln(\mu_C/M_0)| = O(1)$.  The value of $m_{H_u}^2$ at the weak 
scale is then not much smaller than $m_{\tilde{E}}^2$, so that the 
overall scale $M_0$ is not much larger than the weak scale to avoid 
fine-tuning in Eq.~(\ref{eq:ewsb}).  We typically expect $400~{\rm GeV} 
\simlt M_0 \simlt 1~{\rm TeV}$, where the lower bound comes from 
$m_{\tilde{E}} \simgt 100~{\rm GeV}$.  With these values of $M_0$, 
the physical mass for the lightest neutral Higgs boson, $M_{\rm Higgs}$, 
can easily exceed the experimental lower bound of $M_{\rm Higgs} \simgt 
114~{\rm GeV}$~\cite{Barate:2003sz}.  This is because our model provides 
a relatively large value of $A_t$ at the weak scale, so that we can 
avoid the Higgs-boson mass bound with relatively small top squark 
masses.  (The importance of large $A_t$ in reducing fine-tuning was 
particularly emphasized in Refs.~\cite{Kitano:2005wc,Kitano:2006gv}.) 
In Fig.~\ref{fig:Higgs} we plot $M_{\rm Higgs}$, calculated using 
{\it FeynHiggs\,2.4.1}~\cite{Heinemeyer:1998yj}, as a function of $M_0$ 
for $\tan\beta = 5$ (dotted line) and $\tan\beta = 10$ (solid line). 
The $\mu$ parameter is chosen to be $\mu = 150~{\rm GeV}$ arbitrarily, 
but the dependence of the result on $\mu$ is very weak.  From the figure, 
we expect that $M_{\rm Higgs} \simlt 120~{\rm GeV}$.  The value of 
$B$ is given by $B \approx M_0^2/\mu \tan\beta$, so that the preferred 
$\tan\beta$ range of $5 \simlt \tan\beta \simlt 20$ requires a somewhat 
small value of $B$ of order $0.1 M_0^2/\mu$.  The sensitivity of the weak 
scale to variations of the supersymmetric parameters is also not so large 
in this model, since there is no superparticle that has a particularly 
large mass compared with others.  The lightest supersymmetric particle 
(LSP) is either (a neutral component of) the Higgsino or the right-handed 
stau.  In the former case the LSP may be the dark matter of the universe 
if it is produced nonthermally.  In the latter case it will have to 
decay into some neutral particle, e.g. the axino -- the supersymmetric 
partner of the axion, which may compose the dark matter.

Let us now turn to the case that we adopt the ``multiverse,'' or the 
``landscape,'' picture.  More precisely, we now assume that the overall 
supersymmetry breaking parameter $M_0$ has different values in different 
``parts'' of the multiverse, or in different vacua of the theory, with 
larger values preferred by some positive power $n$: $dP \propto dM_0^n$, 
where $P$ is the probability distribution function.  In general the 
distribution of $M_0$ depends on the structure of the supersymmetry 
breaking (uplifting) sector and the sector that produces the constant 
term $c$ in the superpotential, and the assumption of $dP \propto dM_0^n$ 
corresponds typically to tree-level supersymmetry breaking (since 
tree-level supersymmetry breaking naturally prefers larger breaking 
scales).  Note that since environmental selection ``locks'' the value of 
$d$ in Eq.~(\ref{eq:uplift}) as $d \approx |c|^2$, through the condition 
for the cosmological constant being small~\cite{Weinberg:1987dv}, all 
the supersymmetry breaking parameters (including the ones generated 
through direct interactions with the uplifting sector, if any) scale 
in a similar way.  As discussed in Ref.~\cite{Giudice:2006sn}, this 
leads to a small hierarchy between the weak scale and the scale of the 
sfermion masses.  With the statistical pressure of $dP \propto dM_0^n$, 
the sfermion masses $\tilde{m}$ are pushed towards larger values, but not 
beyond the scale where the Higgs mass-squared parameter $m_h^2$ crosses 
zero, since a larger $\tilde{m}$ would lead to the recovery of electroweak 
symmetry in the Higgs sector, a situation hostile to the existence 
of observers.  This leads to $|m_h^2| = O(\tilde{m}^2/8\pi^2) \ll 
\tilde{m}^2$, since $m_h^2$ becomes zero around the scale $\tilde{m}$. 
Moreover, if the parameter $\mu$ scans independently, and if the parameter 
$B$ is sufficiently small at a high scale, then we obtain $|\mu|^2 = 
O(\tilde{m}^2/8\pi^2) \ll \tilde{m}^2$ and thus also $|m_{H_u}^2| = 
O(\tilde{m}^2/8\pi^2) \ll \tilde{m}^2$ --- the property we found in 
our model (see Eq.~(\ref{eq:mHu2})).  In order for this argument to 
be significant, the model must satisfy the conditions for $\mu$ and 
$B$ given above, which we assume to be the case.  An implication of 
this picture is then that the overall scale parameter $M_0$ is expected 
to be somewhat larger than the weak scale: $M_0^2/8\pi^2 \sim |m_{H_u}^2| 
\sim |\mu|^2 \sim v^2$.  The precise hierarchy depends on the strength 
of the pressure $n$, but we generically expect $M_0$ to be in a multi-TeV 
region.  This implies that in this picture all the superparticles, other 
than the Higgsinos, as well as all the $CP$-odd, the heavier $CP$-even, 
and the charged Higgs bosons, $A^0$, $H^0$ and $H^\pm$, have masses in 
this region ($\sim M_0 \simgt 1~{\rm TeV}$).  The ratios between the 
superparticle masses are still given by Eq.~(\ref{eq:prediction-2}), 
and the three Higgs boson masses are given by $m_{A^0} \sim m_{H^0} 
\sim m_{H^\pm} \sim m_{H_d}$.

The spectrum just described can lead to quite distinct phenomenology. 
For example, if $M_0$ is somewhat large, e.g. $M_0 \simgt 2~{\rm TeV}$, 
all the superparticles and heavier Higgs bosons are beyond the discovery 
reach of the LHC, except for the Higgsinos.  Thus, the LHC will 
effectively see the (one Higgs doublet) standard model, plus possibly 
the Higgsinos.  Discoveries of superparticles, however, may be possible 
if $M_0$ is lower.  The LSP is the neutral component of the Higgsinos, 
which may be the dark matter of the universe.  For example, if $M_0 
\simeq 3~{\rm TeV}$, the gravitino mass is $m_{3/2} \simeq 100~{\rm TeV}$, 
and the moduli field mass is $m_T = O(1000\!\sim\!10000~{\rm TeV})$. 
The moduli field is expected to dominate the energy density of the early 
universe, and then it decays into the superparticles and gravitinos, 
which in turn decay into the LSP.  With these masses, the constraint from 
big-bang nucleosynthesis can be avoided (see e.g.~\cite{Asaka:2006bv}) 
and the LSP may compose the dark matter, presumably with some (small) 
amount of dilution of its energy density.  Alternatively, the LSP 
may decay into a lighter particle, e.g. the axino.

\section{Discussions: Predictions from the Landscape?}
\label{sec:concl}

Since it has been difficult to find ways of experimentally ``testing'' 
the landscape picture, it is important to consider what implications 
it can have on the low-energy spectrum and what predictions we can 
get from it when combined with additional assumptions.  In this paper 
we discussed a framework which may either arise from the naturalness 
consideration in the conventional picture or from the landscape picture 
under certain circumstances, and presented an example model which 
leads to strong predictions of the superparticle masses.  The essential 
ingredients of the framework were
\begin{enumerate}
\item[(i)]
The up-type Higgs mass-squared parameter $m_{H_u}^2$ crosses zero 
at a scale close to the superparticle mass scale.
\item[(ii)]
The structure of the theory at the unification (or compactification) 
scale is ``simple'' as far as the observable sector is concerned.
\end{enumerate}
The reason that this can lead to strong predictions, despite the fact 
that each ingredient is not necessarily giving a very strong constraint, 
is that a generic theory satisfying (ii) does not typically lead to 
the property of (i), so that the combination of these two criteria 
can be a very strong constraint on models.  The model we presented 
has a fairly simple structure at the unification scale, arising 
from a simple setup in higher dimensions, and yet gives a vanishing 
$m_{H_u}^2$ at a scale close to the weak scale.  All the supersymmetry 
breaking parameters, except for the holomorphic Higgs mass-squared 
parameter $\mu B$ (and $m_{H_u}^2$), are predicted in terms of a single 
overall mass scale $M_0$ (and $\tan\beta = \langle H_u \rangle/\langle H_d 
\rangle$).  The parameter $M_0$ is expected to be in a multi-TeV region 
if it scans with a preference towards larger values.

On the other hand, it is clear that the specific model discussed 
above is not a unique model satisfying (i) and (ii).  For example, we 
can consider the situation in which the scalar masses are approximately 
universal at a high scale, with somewhat suppressed gaugino masses. 
As observed in~\cite{Feng:1999mn}, this leads to suppressed $m_{H_u}^2$ 
at the weak scale.  In fact, this situation can be realized in the 
setup of Eqs.~(\ref{eq:W},~\ref{eq:uplift}) if the supersymmetry 
breaking (uplifting) sector gives universal scalar masses 
through direct interactions with the observable sector fields.%
\footnote{While completing this paper, we received Ref.~\cite{Abe:2006xp} 
which discusses scenarios with similar spectra, but not in the context 
of the landscape picture, i.e. the picture of scanning parameters.}
In the context of the landscape picture, with a statistical pressure 
acting towards larger values for the supersymmetry breaking masses, 
this can lead to relatively degenerate scalar masses in a multi-TeV 
region and gaugino masses in a few hundred GeV region, with the 
relative gaugino masses still given by Eq.~(\ref{eq:prediction}). 
The Higgsino masses are comparable to the gaugino masses, and 
the value of $\tan\beta$ will be relatively large of $O(10)$, for 
an unsuppressed $B$ parameter.  (The possibility of a relatively 
unsuppressed $B$ parameter is an advantage of unsuppressed scalar 
masses.)  In either of these models, the spectrum of superparticles 
is special such that it leads to a suppressed value of $m_{H_u}^2$ at 
the weak scale, which appears to us as a result of an {\it accidental} 
cancellation due to the specific values of the observed gauge and 
Yukawa couplings.

While the conditions of (i) and (ii) are keys to obtain strong 
predictions for the superparticle spectrum, neither is a necessary 
consequence of the landscape picture.  Indeed, it is possible that 
environmental selection leads to the Higgs mass-squared parameter 
being small due to cancellation {\it between} $m_{H_u}^2$ and $\mu^2$, 
and not just {\it inside} $m_{H_u}^2$, in which case (i) is not 
necessarily satisfied.  Moreover, a simple ultraviolet structure 
may not be preferred by the statistics of landscape vacua, and the 
condition (ii) may also be violated.  In these cases we lose a strong 
constraint on the superparticle spectrum, reducing predictivity, but 
may still get an interesting pattern for the spectrum.  For example, 
landscape statistics may prefer the case in which the supersymmetry 
breaking (uplifting) sector gives somewhat {\it random} scalar masses 
through direct interactions with the observable sector, in the setup 
of Eqs.~(\ref{eq:W},~\ref{eq:uplift}).  (Flavor universality may 
still have to be assumed unless these masses are very large.) 
With the statistical pressure of pushing the overall mass scale 
to larger values, we find the Higgs mass-squared parameter somewhat 
suppressed compared with the scalar superparticle masses.  The 
spectrum will then contain the scalar superparticles and the Higgsinos 
in a multi-TeV region, whose masses do not obey simple relations. 
The gaugino masses, however, may still be of order a few hundred GeV 
and obey Eq.~(\ref{eq:prediction}) in the case that the direct effect 
from the supersymmetry breaking sector is suppressed in the gauge kinetic 
functions.  Deviations from Eq.~(\ref{eq:prediction}), however, can 
also occur, e.g., if the moduli-stabilization and uplifting sectors 
deviate from the minimal form of Eqs.~(\ref{eq:W},~\ref{eq:uplift}), 
in which case the gaugino masses unify at a scale that is not necessarily 
the intermediate scale of Eq.~(\ref{eq:M_mess}), or if the direct 
effect is not suppressed in the gauge kinetic functions, in which 
case the gaugino masses are of order a multi-TeV.  The value of 
$\tan\beta$ will generically be of $O(1)$ for an unsuppressed value 
for the $B$ parameter.

To summarize, we have argued that both the conventional naturalness 
picture and the landscape picture (with certain assumptions) may point 
to a scenario in which $m_{H_u}^2$ crosses zero near the weak scale. 
Combining this constraint with a simple ultraviolet structure can lead 
to a highly predictive superparticle spectrum, an example being the 
model presented in section~\ref{sec:model}.  The model predicts all 
the supersymmetry breaking parameters, except for the holomorphic 
Higgs mass-squared parameter $\mu B$ (and $m_{H_u}^2$), in terms of 
a single overall mass scale $M_0$ (with a weak dependence on $\tan\beta$). 
This parameter is expected to be of order a few hundred GeV if it does 
not scan but in a multi-TeV region if it does scan, allowing for the 
possibility of experimentally distinguishing between these two cases. 
We have also discussed implications of the landscape picture on 
the supersymmetry breaking masses in a general setup arising from 
Eqs.~(\ref{eq:W},~\ref{eq:uplift}) with possible additional interactions. 
Depending on the form of these interactions, strong predictions on the 
entire superparticle masses may be lost, but some predictions, such as 
those on the gaugino masses, may still be preserved.  It is interesting 
that the experimental observation of one of these spectra may hint at 
possible statistical pressures acting on parameters of the theory, and 
thus the gross structure of vacua in the ``vicinity'' of our own one.

\section*{Acknowledgments}

We thank Lawrence Hall for discussions.  This work was supported 
in part by the Director, Office of Science, Office of High Energy 
and Nuclear Physics, of the US Department of Energy under Contract 
DE-AC02-05CH11231.  The work of Y.N. was also supported by the National 
Science Foundation under grant PHY-0555661, by a DOE Outstanding Junior 
Investigator award, and by an Alfred P. Sloan Research Fellowship.


\newpage


\begin{thebibliography}{99}

\bibitem{Dimopoulos:1981zb}
S.~Dimopoulos and H.~Georgi,
%``Softly Broken Supersymmetry And SU(5),''
Nucl.\ Phys.\ B {\bf 193}, 150 (1981);
%%CITATION = NUPHA,B193,150;%%
%\bibitem{Sakai:1981gr}
N.~Sakai,
%``Naturalness In Supersymmetric 'Guts',''
Z.\ Phys.\ C {\bf 11}, 153 (1981);
%%CITATION = ZEPYA,C11,153;%%
%\bibitem{Dimopoulos:1981yj}
S.~Dimopoulos, S.~Raby and F.~Wilczek,
%``Supersymmetry And The Scale Of Unification,''
Phys.\ Rev.\ D {\bf 24}, 1681 (1981).
%%CITATION = PHRVA,D24,1681;%%

\bibitem{Kitano:2006gv}
R.~Kitano and Y.~Nomura,
%``Supersymmetry, naturalness, and signatures at the LHC,''
Phys.\ Rev.\ D {\bf 73}, 095004 (2006)
[arXiv:hep-ph/0602096].
%%CITATION = HEP-PH 0602096;%%

\bibitem{Feng:1999mn}
J.~L.~Feng, K.~T.~Matchev and T.~Moroi,
%``Multi-TeV scalars are natural in minimal supergravity,''
Phys.\ Rev.\ Lett.\  {\bf 84}, 2322 (2000)
[arXiv:hep-ph/9908309];
%%CITATION = HEP-PH 9908309;%%
%\bibitem{Feng:1999zg}
%J.~L.~Feng, K.~T.~Matchev and T.~Moroi,
%``Focus points and naturalness in supersymmetry,''
Phys.\ Rev.\ D {\bf 61}, 075005 (2000)
[arXiv:hep-ph/9909334].
%%CITATION = HEP-PH 9909334;%%

\bibitem{Choi:2005hd}
K.~Choi, K.~S.~Jeong, T.~Kobayashi and K.~i.~Okumura,
%``Little SUSY hierarchy in mixed modulus-anomaly mediation,''
Phys.\ Lett.\ B {\bf 633}, 355 (2006)
[arXiv:hep-ph/0508029].
%%CITATION = HEP-PH 0508029;%%

\bibitem{Kitano:2005wc}
R.~Kitano and Y.~Nomura,
%``A solution to the supersymmetric fine-tuning problem within the MSSM,''
Phys.\ Lett.\ B {\bf 631}, 58 (2005)
[arXiv:hep-ph/0509039].
%%CITATION = HEP-PH 0509039;%%

\bibitem{Weinberg:1987dv}
S.~Weinberg,
%``ANTHROPIC BOUND ON THE COSMOLOGICAL CONSTANT,''
Phys.\ Rev.\ Lett.\  {\bf 59}, 2607 (1987);
%%CITATION = PRLTA,59,2607;%%
%\bibitem{Martel:1997vi}
H.~Martel, P.~R.~Shapiro and S.~Weinberg,
%``Likely Values of the Cosmological Constant,''
Astrophys.\ J.\  {\bf 492}, 29 (1998)
[arXiv:astro-ph/9701099].
%%CITATION = ASTRO-PH 9701099;%%

\bibitem{Bousso:2000xa}
R.~Bousso and J.~Polchinski,
%``Quantization of four-form fluxes and dynamical neutralization of the
%cosmological constant,''
JHEP {\bf 0006}, 006 (2000)
[arXiv:hep-th/0004134];
%%CITATION = HEP-TH 0004134;%%
%\bibitem{Susskind:2003kw}
L.~Susskind,
%``The anthropic landscape of string theory,''
arXiv:hep-th/0302219;
%%CITATION = HEP-TH 0302219;%%
%\bibitem{Douglas:2003um}
M.~R.~Douglas,
%``The statistics of string / M theory vacua,''
JHEP {\bf 0305}, 046 (2003)
[arXiv:hep-th/0303194];
%%CITATION = HEP-TH 0303194;%%
and references therein.

\bibitem{Giudice:2006sn}
G.~F.~Giudice and R.~Rattazzi,
%``Living dangerously with low-energy supersymmetry,''
Nucl.\ Phys.\ B {\bf 757}, 19 (2006)
[arXiv:hep-ph/0606105].
%%CITATION = HEP-PH 0606105;%%

\bibitem{Arkani-Hamed:2004fb}
N.~Arkani-Hamed and S.~Dimopoulos,
%``Supersymmetric unification without low energy supersymmetry and  signatures
%for fine-tuning at the LHC,''
JHEP {\bf 0506}, 073 (2005)
[arXiv:hep-th/0405159];
%%CITATION = HEP-TH 0405159;%%
%\bibitem{Giudice:2004tc}
G.~F.~Giudice and A.~Romanino,
%``Split supersymmetry,''
Nucl.\ Phys.\ B {\bf 699}, 65 (2004)
[Erratum-ibid.\ B {\bf 706}, 65 (2005)]
[arXiv:hep-ph/0406088].
%%CITATION = HEP-PH 0406088;%%

\bibitem{Feldstein:2006ce}
See, e.g., 
B.~Feldstein, L.~J.~Hall and T.~Watari,
%``Landscape prediction for the Higgs boson and top quark masses,''
arXiv:hep-ph/0608121.
%%CITATION = HEP-PH 0608121;%%

\bibitem{Chamseddine:1982jx}
A.~H.~Chamseddine, R.~Arnowitt and P.~Nath,
%``Locally Supersymmetric Grand Unification,''
Phys.\ Rev.\ Lett.\  {\bf 49}, 970 (1982);
%%CITATION = PRLTA,49,970;%%
%\bibitem{Barbieri:1982eh}
R.~Barbieri, S.~Ferrara and C.~A.~Savoy,
%``Gauge Models With Spontaneously Broken Local Supersymmetry,''
Phys.\ Lett.\ B {\bf 119}, 343 (1982);
%%CITATION = PHLTA,B119,343;%%
%\bibitem{Hall:1983iz}
L.~J.~Hall, J.~Lykken and S.~Weinberg,
%``Supergravity As The Messenger Of Supersymmetry Breaking,''
Phys.\ Rev.\ D {\bf 27}, 2359 (1983).
%%CITATION = PHRVA,D27,2359;%%

\bibitem{Agrawal:1997gf}
V.~Agrawal, S.~M.~Barr, J.~F.~Donoghue and D.~Seckel,
%``The anthropic principle and the mass scale of the standard model,''
Phys.\ Rev.\ D {\bf 57}, 5480 (1998)
[arXiv:hep-ph/9707380].
%%CITATION = HEP-PH 9707380;%%

\bibitem{Kawamura:2000ev}
Y.~Kawamura,
%``Triplet-doublet splitting, proton stability and extra dimension,''
Prog.\ Theor.\ Phys.\  {\bf 105}, 999 (2001)
[arXiv:hep-ph/0012125];
%%CITATION = HEP-PH 0012125;%%
%\bibitem{Hall:2001pg}
L.~J.~Hall and Y.~Nomura,
%``Gauge unification in higher dimensions,''
Phys.\ Rev.\ D {\bf 64}, 055003 (2001)
[arXiv:hep-ph/0103125];
%%CITATION = HEP-PH 0103125;%%
%\bibitem{Hall:2002ea}
%L.~J.~Hall and Y.~Nomura,
%``Grand unification in higher dimensions,''
Annals Phys.\  {\bf 306}, 132 (2003)
[arXiv:hep-ph/0212134].
%%CITATION = HEP-PH 0212134;%%

\bibitem{Kachru:2003aw}
S.~Kachru, R.~Kallosh, A.~Linde and S.~P.~Trivedi,
%``De Sitter vacua in string theory,''
Phys.\ Rev.\ D {\bf 68}, 046005 (2003)
[arXiv:hep-th/0301240].
%%CITATION = HEP-TH 0301240;%%

\bibitem{Choi:2004sx}
K.~Choi, A.~Falkowski, H.~P.~Nilles, M.~Olechowski and S.~Pokorski,
%``Stability of flux compactifications and the pattern of supersymmetry
%breaking,''
JHEP {\bf 0411}, 076 (2004)
[arXiv:hep-th/0411066];
%%CITATION = HEP-TH 0411066;%%
%\bibitem{Choi:2005ge}
K.~Choi, A.~Falkowski, H.~P.~Nilles and M.~Olechowski,
%``Soft supersymmetry breaking in KKLT flux compactification,''
Nucl.\ Phys.\ B {\bf 718}, 113 (2005)
[arXiv:hep-th/0503216].
%%CITATION = HEP-TH 0503216;%%

\bibitem{Choi:2005uz}
K.~Choi, K.~S.~Jeong and K.~i.~Okumura,
%``Phenomenology of mixed modulus-anomaly mediation in fluxed string
%compactifications and brane models,''
JHEP {\bf 0509}, 039 (2005)
[arXiv:hep-ph/0504037].
%%CITATION = HEP-PH 0504037;%%

\bibitem{Skands:2003cj}
P.~Skands {\it et al.},
%``SUSY Les Houches accord: Interfacing SUSY spectrum calculators, decay
%packages, and event generators,''
JHEP {\bf 0407}, 036 (2004)
[arXiv:hep-ph/0311123].
%%CITATION = HEP-PH 0311123;%%
%\verb|http://home.fnal.gov/~skands/slha|

\bibitem{Barbieri:2001yz}
R.~Barbieri, L.~J.~Hall and Y.~Nomura,
%``Softly broken supersymmetric desert from orbifold compactification,''
Phys.\ Rev.\ D {\bf 66}, 045025 (2002)
[arXiv:hep-ph/0106190];
%%CITATION = HEP-PH 0106190;%%
%\bibitem{Barbieri:2001dm}
%R.~Barbieri, L.~J.~Hall and Y.~Nomura,
%``Models of Scherk-Schwarz symmetry breaking in 5D: Classification and
%calculability,''
Nucl.\ Phys.\ B {\bf 624}, 63 (2002)
[arXiv:hep-th/0107004].
%%CITATION = HEP-TH 0107004;%%

\bibitem{Brubaker:2006xn}
E.~Brubaker {\it et al.}  [Tevatron Electroweak Working Group],
%``Combination of CDF and D0 results on the mass of the top quark,''
arXiv:hep-ex/0608032.
%%CITATION = HEP-EX 0608032;%%

\bibitem{Barate:2003sz}
R.~Barate {\it et al.}  [ALEPH Collaboration],
%``Search for the standard model Higgs boson at LEP,''
Phys.\ Lett.\ B {\bf 565}, 61 (2003)
[arXiv:hep-ex/0306033];
%%CITATION = HEP-EX 0306033;%%
% \bibitem{unknown:2001xx}
LEP Higgs Working Group Collaboration,
%``Searches for the neutral Higgs bosons of the MSSM: Preliminary combined
%results using LEP data collected at energies up to 209-GeV,''
arXiv:hep-ex/0107030.
%%CITATION = HEP-EX 0107030;%%

\bibitem{Heinemeyer:1998yj}
S.~Heinemeyer, W.~Hollik and G.~Weiglein,
%``FeynHiggs: A program for the calculation of the masses of the neutral
%CP-even Higgs bosons in the MSSM,''
Comput.\ Phys.\ Commun.\  {\bf 124}, 76 (2000)
[arXiv:hep-ph/9812320];
%%CITATION = HEP-PH 9812320;%%
%\bibitem{Heinemeyer:1998np}
%S.~Heinemeyer, W.~Hollik and G.~Weiglein,
%``The masses of the neutral CP-even Higgs bosons in the MSSM: Accurate
%analysis at the two-loop level,''
Eur.\ Phys.\ J.\ C {\bf 9}, 343 (1999)
[arXiv:hep-ph/9812472].
%%CITATION = HEP-PH 9812472;%%

\bibitem{Asaka:2006bv}
T.~Asaka, S.~Nakamura and M.~Yamaguchi,
%``Gravitinos from heavy scalar decay,''
Phys.\ Rev.\ D {\bf 74}, 023520 (2006)
[arXiv:hep-ph/0604132];
%%CITATION = HEP-PH 0604132;%%
M.~Endo, K.~Hamaguchi and F.~Takahashi,
%``Moduli-induced gravitino problem,''
Phys.\ Rev.\ Lett.\  {\bf 96}, 211301 (2006)
[arXiv:hep-ph/0602061].
%%CITATION = HEP-PH 0602061;%%

\bibitem{Abe:2006xp}
H.~Abe, T.~Higaki, T.~Kobayashi and Y.~Omura,
%``Moduli stabilization, F-term uplifting and soft supersymmetry breaking
%terms,''
arXiv:hep-th/0611024.
%%CITATION = HEP-TH 0611024;%%

\end{thebibliography}
\end{document}